\documentclass[aps,prl,twocolumn,groupedaddress,showpacs]{revtex4}

\usepackage{graphics}

\begin{document}

\title{Universal properties in ultracold ion-atom interactions}


\author{Bo Gao}
\affiliation{Department of Physics and Astronomy,
	University of Toledo, MS 111,
	Toledo, Ohio 43606}

\date{January 26, 2010}

\begin{abstract}

We present some of the universal properties in ion-atom
interaction derived from a newly formulated quantum-defect theory
for $-1/r^4$ type of long-range interactions.
For bound states, we present the
universal bound spectrum, namely the equivalent of the Rydberg
formula, for ion-atom systems. 
For scattering, we introduce the concept of universal 
resonance spectrum to give a systematic understanding
of many resonances present in ion-atom scattering. 
The theory further provides a method for an accurate 
spectroscopic determination of the atomic polarizability.
It also suggests the existence of atom-like
molecules, in which multiple atoms orbit around a heavy
ion.

\end{abstract}


\maketitle

As experimental techniques for preparing and manipulating
cold atomic ions improve \cite{gri09},
there is a rapidly growing interest in ion-atom interactions at
ultracold temperatures \cite{cot00,cot02,idz09,zha09}.
While it is intuitively clear that an ion can exert a much stronger influence
on its environment than either an atom or a molecule, 
the precise nature of this influence,
especially at the quantum level, is far from being understood.
For example, an ion-atom system is one of the few types of 
two-body systems for which
their loosely-bound states, namely the highly excited
rovibrational states right below the dissociation limit,
are yet to be observed experimentally or fully characterized
theoretically \cite{raa09}. 

This work presents part of a newly formulated
quantum-defect theory (QDT) for ion-atom interactions.
It gives precise explanation and characterization
of the meaning and the consequence of the ``strong
influence'' of an ion on a neighboring atom.
On a more technical level, it is a version of the QDT
for $-1/r^4$ type of long-range potential \cite{oma61,wat80,fab86}
that brings our understanding of ion-atom interactions 
to the same level as, and in certain areas
exceeding, our current understanding of ultracold 
atom-atom interactions \cite{chi10}. 
The theory is a result of combining conceptual 
developments, built further upon Refs.~\cite{gao01,gao00,gao08a},
with improved mathematical understanding of the modified Mathieu
functions \cite{abr64,hol73,khr93},
especially for negative energies.
The latter development, which allows for the efficient
determination of all QDT functions for $-1/r^4$ potential,
is achieved by solving the modified Mathieu equation
using techniques we have previously developed for
the analytic solutions of $1/r^6$ \cite{gao98a} and 
$1/r^3$ \cite{gao99a} potentials. 

We focus here on the universal spectrum for ion-atom interactions.
For bound states, it is the equivalent of the 
Rydberg formula for $-1/r^4$ type of long-range potential,
formulated in a way to take advantage of
the angular momentum insensitivity of the short-range parameter 
 that is a characteristic of atom-atom
and ion-atom interactions \cite{gao01,gao04b,gao08a}.
It allows for the determination of the entire rovibrational spectrum 
in the threshold region from a single short-range parameter, 
such as the quantum defect. More importantly,
the concept of universal spectrum is generalized
here to positive energies to include 
scattering resonance positions.
It facilitates a systematic understanding of many resonances
that are present in ion-atom scattering \cite{cot00,idz09,zha09}, 
within the same framework, and using the same short-range 
parameter, as the bound spectrum.
A number of more specific results will be extracted out of
the universal spectrum, including the energy bins \cite{gao00,chi10}
that contain various loosely-bound states or scattering resonances. 
Furthermore, we show that there exists a critical energy,
below which all scattering resonances have positive widths
that correspond to time delay \cite{wig55}, and above which all scattering
resonances have negative widths that correspond to
time advance \cite{wig55}. Such concepts, together with the 
concept of universal width function
to be briefly discussed there, provide what we believe to be the
first systematic understanding of the global structure of 
scattering. Conceptually, they are expected to 
be generally applicable to
all quantum systems with $-1/r^n$ type of long-range
potential with $n>2$.
We briefly outline selective qualitative features of ion-atom
interaction before getting to the mathematical formulation 
and the results of the theory.

Like other quantum systems around the threshold,
ion-atom interaction in the ultracold regime is determined
primarily by its long-range interaction. Assuming that the atom
involved has no permanent quadrupole, such as the case for
alkali-metal atoms in their ground states, the long-range
ion-atom interaction has the form of
$V(r)\sim -C_4/r^4$
where $C_4$ is related, in atomic units, to the static
dipole polarizability of the atom, $\alpha_1$, by $C_4=\alpha_1/2$.
The qualitative features of ion-atom interaction can be
understood through the length scale, $\beta_4=(2\mu C_4/\hbar^2)^{1/2}$,
and other related scale parameters that are associated with the 
long-range potential.
\begin{table}
\caption{Sample scale parameters for ion-atom systems. 
$C_4$, in atomic units, is given by $\alpha_1/2$ where $\alpha_1$ is the 
static dipole polarizability of the atom.
The $\beta_4=(2\mu C_4/\hbar^2)^{1/2}$ is the length scale. 
The $s_E=(\hbar^2/2\mu)(1/\beta_4)^2$ 
is the corresponding energy scale. It is given both in units of microkelvin and in
units of kHz.\label{tb:scales}}
\begin{ruledtabular}
\begin{tabular}{lrrrrr}
System                    & $C_4$ (a.u.)          & $\beta_4$ (a.u.) & $s_E/k_B$ ($\mu$K) & $s_E/h$ (kHz) \\
\hline
$^{87}$Rb$^{+}$-$^6$Li    & 82.06\footnotemark[1] & $1297$           & $9.148$            & $190.6$       \\
$^{87}$Rb$^+$-$^{23}$Na   & 81.30\footnotemark[2] & $2321$           & $0.8841$           & $18.42$       \\
$^6$Li$^+$-$^6$Li         & 82.06\footnotemark[1] & $948.5$          & $32.01$            & $667.0$       \\
$^{23}$Na$^+$-$^{23}$Na   & 81.30\footnotemark[2] & $1846$           & $2.212$            & $46.08$       \\
$^{40}$K$^+$-$^{40}$K     & 145.1\footnotemark[2] & $3251$           & $0.4101$           & $8.545$       \\
$^{133}$Cs$^+$-$^{133}$Cs & 199.9\footnotemark[2] & $6959$           & $0.02691$          & $0.5608$      \\
\end{tabular}
\end{ruledtabular}
\footnotetext[1]{Using $\alpha_1$ from Ref.~\cite{yan96}.}
\footnotetext[2]{Using $\alpha_1$ from Ref.~\cite{der99}.}
\end{table}
Table~\ref{tb:scales} gives samples of such parameters for alkali-metal 
systems. The length scale
$\beta_4$ is a measure of the size of the last few loosely-bound states.
The corresponding energy scale, $s_E=(\hbar^2/2\mu)(1/\beta_4)^2$, sets the
scale for their energy spacing. The related time scale $s_T=\hbar/s_E$ sets
the scale for, e.g., the lifetime of a shape resonance, and
$\pi\beta_4^2$ sets the scale for low-energy scattering cross section.

A large length scale $\beta_4$ is what characterizes the 
``strong influence'' of an ion on its environment.
From Table~\ref{tb:scales}, it is clear that ion-atom interaction
has a much longer length scale than either the atom-atom interaction,
with a corresponding length scale of $\beta_6\sim 100$ a.u., or
the electron-atom interaction, with a corresponding length scale 
of $\beta_4\sim 10$ a.u.
More importantly, the $\beta_4$ for ion-atom 
is hundreds times greater than $r_0$, the distance
below which the ion-atom interaction starts to deviate
from the $-C_4/r^4$ behavior, and which can be estimated to be
of the order of $r_0\sim 10$-30 a.u.. 
The implication of this large scale separation is that an 
ion-atom system typically supports thousands of 
bound states, and, from Levinson theorem \cite{lev49}, thousands of 
scattering resonances,
with many of them around the threshold following the universal behavior 
as described by the angular-momentum-insensitive formulation of the 
quantum-defect theory (AQDT) \cite{gao01,gao08a}.

\textit{Universal bound spectrum}:
In AQDT \cite{gao01,gao08a}, 
the bound spectrum of a two-body system with 
$-1/r^n$ ($n>2$) type of long-range potential is given rigorously
by the solutions of  
\begin{equation}
\chi^{c(n)}_l(\epsilon_s) = K^c(\epsilon,l) \;.
\label{eq:ubsp}
\end{equation}
Here the $\chi^{c(n)}_l$ function is a universal
function of a scaled energy, $\epsilon_s=\epsilon/s_E$, 
that is uniquely determined by the exponent of the 
long-range interaction $n$ and the angular momentum $l$. 
All the short-range physics is encapsulated in the parameter 
$K^c(\epsilon,l)$. It is a short-range $K$ matrix that is
a meromorphic function of both $\epsilon$
and $l$ \cite{gao08a}.
The $K^c$ parameter is further related to the quantum defect 
$\mu^c$, which is defined to have a 
range of $0\le \mu^c<1$, by
$
K^c(\epsilon,l) = \tan[\pi\mu^c(\epsilon,l)+\pi b/2]
$
where $b=1/(n-2)$ \cite{gao08a}.

For $n=4$, we have obtained all QDT functions of Ref.~\cite{gao08a}
through an improved understanding of the modified Mathieu 
equation \cite{abr64,hol73,khr93}.
In particular, we have obtained the universal
$\chi^{c(n)}_l$ function for $n=4$:
\begin{equation}
\chi^{c(4)}_l = \tan(\pi\nu/2)\frac{1+M_{\epsilon_s l}^2}{1-M_{\epsilon_s l}^2} \;,
\label{eq:chic4}
\end{equation}
where $\nu$ is the characteristic exponent for $-1/r^4$ type of 
potential \cite{abr64,hol73,khr93,sup}, and
\begin{eqnarray}
M_{\epsilon_s l}(\nu) &=& 2^{-2\nu}|\epsilon_s|^{\nu/2}
	\left(\frac{\Gamma[1-(\nu+\nu_0)/2]}{\Gamma[1+(\nu+\nu_0)/2)]}\right) \nonumber\\
	& &\times \left(\frac{\Gamma[1-(\nu-\nu_0)/2]}{\Gamma[1+(\nu-\nu_0)/2]}\right)
	\left(\frac{C_{\epsilon_s l}(-\nu)}{C_{\epsilon_s l}(+\nu)}\right) \;,
\label{eq:Mnu4}
\end{eqnarray}
in which $\nu_0=l+1/2$,
\begin{equation}
C_{\epsilon_s l}(\nu) = \prod_{j=0}^{\infty} Q(\nu+2j) \;,
\label{eq:Cnu4}
\end{equation}
with $Q(\nu)$ being given by a continued fraction
\begin{equation}
Q(\nu) = \frac{1}{1-\frac{\epsilon_s}{
	[(\nu+2)^2-\nu_0^2][(\nu+4)^2-\nu_0^2]}
	Q(\nu+2)} \;.
\label{eq:Qcf4}
\end{equation}

\begin{figure}[ht]
\scalebox{0.3}{\includegraphics{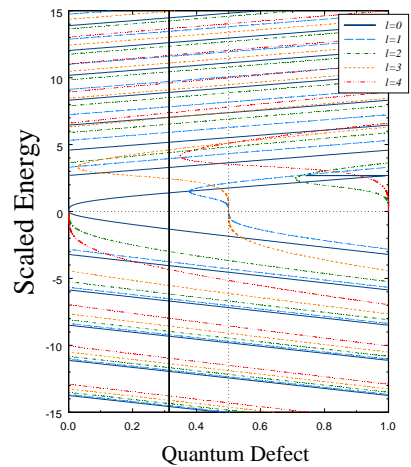}}
\caption{(Color online) The universal spectrum for two-body quantum systems with 
$-1/r^4$ type of interactions, including both the universal
bound spectrum for $\epsilon<0$, where the y-axis is
defined as $-(-\epsilon_s)^{1/4}$, and the universal resonance spectrum for
$\epsilon>0$, where the y-axis is defined as $(\epsilon_s)^{1/4}$. 
For any two-body quantum system with 
$-1/r^4$ type of long-range potential, the bound spectrum
and the resonance spectrum are given by the cross points between
the curves plotted and the curves representing
$\mu^c(\epsilon,l)$. For systems such as ion-atom,
for which $\beta_4\gg r_0$, $\mu^c(\epsilon,l)$ is approximately
an $l$-independent constant, allowing the determination of
the entire rovibrational spectrum, and the resonance spectrum,
from a single parameter, as illustrated by a vertical line in
the figure.
\label{fig:usp4}}
\end{figure}
Equations~(\ref{eq:ubsp}) and (\ref{eq:chic4}) give 
the universal bound spectrum for $-1/r^4$ type
of long-range interaction.
Figure~\ref{fig:usp4} gives one of its possible graphical
representations, in combination with the universal resonance
spectrum to be discussed next.
\begin{table}
\caption{Energy bins for the first few partial waves.
The $i$-th bound state of angular momentum $l$, with $i=1$ 
corresponding to the least-bound state, is to be
found within $B_{-i}\le\epsilon_s< B_{-i+1}$ for $i>1$,
and within $B_{-1}\le \epsilon_s< 0$ for $i=1$.
Shape resonances of angular momentum $l$ can only exist
between $0<\epsilon_{s}<B_c$. 
The $i$-th diffraction resonance of angular momentum $l$,
defined as a resonance with negative width, 
is to be found within $B_{i-1}<\epsilon_s\le B_i$ for
$i\ge 1$. A zeroth diffraction resonance may exist 
within $B_c< \epsilon_s\le B_{0}$, depending on the
quantum defect.
\label{tb:ebins}}
\begin{ruledtabular}
\begin{tabular}{lrrrrr}
$B_x$    & $l=0$   & $l=1$   & $l=2$   & $l=3$   & $l=4$   \\
\hline
$B_{4}$  & 4782.3  & 10266   & 19073   & 32206   & 50780   \\
$B_{3}$  & 1734.9  & 4480.1  & 9371.7  & 17218   & 28940   \\
$B_{2}$  & 442.02  & 1553.1  & 3886.6  & 8057.0	 & 14792   \\
$B_{1}$  & 52.766  & 350.37  & 1199.8  & 3021.3  & 6348.0  \\
$B_{0}$  & 0       & 20.514  & 174.75  & 687.38  & 1896.0  \\
$B_{c}$  & 0       & 4.8878  & 35.505  & 120.16	 & 292.38  \\
\hline
$B_{-1}$ & -105.81 & -336.54 & -753.97 & -1407.4 & -2345.8 \\
$B_{-2}$ & -1179.9 & -2432.9 & -4284.6 & -6835.1 & -10184  \\
$B_{-3}$ & -5207.5 & -8840.6 & -13708  & -19961  & -27750  \\
$B_{-4}$ & -15308  & -23247  & -33279  & -45607  & -60433  
\end{tabular}										
\end{ruledtabular}									
\end{table}

\textit{Universal scattering properties}:
In AQDT, the single-channel scattering properties are
determined from \cite{gao08a}
\begin{equation}
K_l \equiv \tan\delta_l = 
	\left(Z^{c(n)}_{gc}K^c -Z^{c(n)}_{fc}\right)
	\left(Z^{c(n)}_{fs}-Z^{c(n)}_{gs}K^c \right)^{-1} \;,
\label{eq:qdtpe}
\end{equation}
where $Z^{c(n)}_{xy}(\epsilon_s,l)$ are again universal
QDT functions that are uniquely determined by the long-range
exponent $n$ and the angular momentum $l$ \cite{gao08a}.

We use here the definition of scattering resonance positions 
as energies at which $\sin^2\delta_l=1$, namely the energies 
at which the scattering cross section reaches its 
unitarity limit \cite{sup}.
Such locations 
can be determined as the roots of the denominator in
Eq.~(\ref{eq:qdtpe}).
Defining a generalized $\chi^{c(n)}_l$ function
for positive energies as
$
\widetilde{\chi}^{c(n)}_l(\epsilon_s)\equiv Z^{c(n)}_{fs}/Z^{c(n)}_{gs}
$,
the resonance positions can be formulated in a similar
fashion as the bound spectrum, as the solutions of
\begin{equation}
\widetilde{\chi}^{c(n)}_l(\epsilon_s) = K^c(\epsilon,l) \;.
\label{eq:ursp}
\end{equation}
For $n=4$, we have obtained
\begin{equation}
\widetilde{\chi}^{c(4)}_l = \tan(\pi\nu/2)
	\frac{1-(-1)^lM_{\epsilon_s l}^2\tan[\pi(\nu-\nu_0)/2]}
	{1+(-1)^lM_{\epsilon_s l}^2\tan[\pi(\nu-\nu_0)/2]} \;,
\label{eq:gchic4}
\end{equation}
in which $\nu$, $\nu_0$, and $M_{\epsilon_s l}$ are the same
as those defined earlier in relation to the bound spectrum.

Equations~(\ref{eq:ursp}) and (\ref{eq:gchic4}) give 
the universal resonance spectrum for $-1/r^4$ type
of long-range interaction. The $\epsilon>0$ portion of
Fig.~\ref{fig:usp4} gives one of its possible graphical 
representations. Similar to a bound spectrum, which
describes, over a set of discrete energies, the rise of a 
phase from zero to a finite value
at the threshold, a resonance spectrum describes its 
subsequent evolution back towards zero.
For a scattering resonance located at $\epsilon_{sl}$,
which is one of the solutions of Eq.~(\ref{eq:ursp}),
the width of the resonance, again derived from Eq.~(\ref{eq:qdtpe}), 
is given by
\begin{equation}
\gamma_l(\epsilon_{sl}) = -2\left\{\left.\frac{d\widetilde{\chi}^{c(n)}_l}{d\epsilon_s}
	\right|_{\epsilon_{sl}}\left[Z^{c(n)}_{gs}(\epsilon_{sl},l)\right]^2\right\}^{-1} \;.
\label{eq:gammal}
\end{equation}	
It shows that while the position of a scattering resonance
depends on the short-range parameter,
the width of the resonance, as a function of the resonance
position, follows a universal behavior that is uniquely 
determined by the long-range interaction.
We call Eq.~(\ref{eq:gammal}) the universal width function.

\begin{figure}
\scalebox{0.35}{\includegraphics{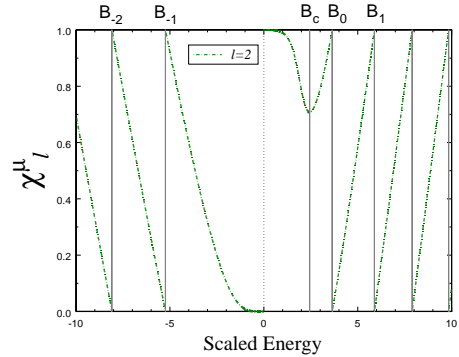}}
\caption{(Color online) Illustrating definitions of energy bins, and the critical
energy $B_c$, using the case $d$ wave. The ``Scaled Energy'' is defined
as in Figure~\ref{fig:usp4}. The $y$-axis is defined as
$\chi^\mu_l = [\tan^{-1}(\chi^c_l)-\pi b/2]/\pi$, in which
$\tan^{-1}(\chi^c_l)$ is taken to be within a range of $\pi$
of $[\pi b/2, \pi+\pi b/2)$. Figure~\ref{fig:usp4} is made up of
multiple such curves for different $l$, except that they are turned
around for a better visualization of the spectrum.
\label{Figure2}}
\end{figure}
There are a number of conclusions 
that can be drawn from the universal spectrum.
We give here a brief summary, with more complete
discussions to be given elsewhere.
(a) For ion-atom systems, namely systems with $\beta_4\gg r_0$, 
the universal spectrum allows the determination of
the entire rovibrational spectrum and the entire resonance
spectrum in the threshold region, including states of
different $l$, using a single parameter,
assuming $\alpha_1$ is known accurately. This single parameter
can be determined from the measurement of either a single
bound state energy \cite{gao01}, or a single resonance position.
(b) From measurements of two or more bound state
energies and/or resonance positions, and using the method
of Ref.~\cite{gao01} and its generalizations, the universal
spectrum can give not only an accurate prediction of the entire
spectrum, but also an accurate, spectroscopic determination of
the atomic dipole polarizability, with no knowledge required
of the details of the atomic interaction at short 
distances \cite{gao01}. The accuracy that can be expected with 
such a measurement is conservatively estimated to be $10^{-6}$,
far better than the best achieved using atomic
interferometry \cite{cro09}.
(c) The measurements of bound state energies and resonance
positions are further facilitated by the energy bin 
concept \cite{gao00,chi10} embedded in the universal
spectrum. They give ranges of energies in which a certain
state is to be found, as explained and tabulated in 
Table~\ref{tb:ebins} and illustrated in Fig.~\ref{Figure2}.
The earlier version of the 
concept \cite{gao00,chi10} has been generalized here to include
bins for resonance states.
(d) The properties of the $\widetilde{\chi}^c_l$ functions show that 
there exists a critical energy for each $l$, $B_c(l)$, defined by
$\left.d\widetilde{\chi}^{c(n)}_l/d\epsilon_s\right|_{B_c(l)}=0$. 
Below $B_c(l)$, $\widetilde{\chi}^{c(n)}_l$ is a
piecewise monotonically decreasing function of energy with
$d\widetilde{\chi}^{c(n)}_l/d\epsilon_s<0$, 
which,
from Eq.~(\ref{eq:gammal}), implies that all resonances occurring in
this region have positive widths corresponding to time delays \cite{wig55}.
Above $B_c(l)$, $\widetilde{\chi}^{c(n)}_l$ evolves into a
piecewise monotonically increasing function of energy with
$d\widetilde{\chi}^{c(n)}_l/d\epsilon_s>0$, 
implying that all resonances above $B_c(l)$ have 
negative widths corresponding to time advances \cite{wig55}.
Such resonances will be called diffraction resonances \cite{sup}
to distinguish them from shape resonances, which have positive widths.
We note that the incorporation of diffraction resonances
into the resonance spectrum is essential to complete 
the physical picture of the evolution of the phase back towards zero.
(e) The universal spectrum gives the maximum number of shape resonances
that can exist for a particular $l$. As can be observed from
Figure~\ref{fig:usp4}, this number is one for $l=1$ through 4.
Results for other $l$ will be presented elsewhere, which show,
e.g., that the minimum $l$ that can support two shape resonances
is $l=7$ for ion-atom interactions.
(f) The results here provide explicit illustrations and confirmations 
of the general properties of $-1/r^n$ ($n>2$) type of interactions derived 
earlier, such as the breakdown of the semiclassical approximation around
the threshold (see Ref.~\cite{gao08a} and the references therein), 
and the conclusion that the least-bound state 
for an ion-atom system is either an $s$ state or a $p$ state,
depending on the quantum defect \cite{gao04b}.
(g) The large size of an ion-atom bound state, 
as compared to the range of atom-atom interaction, 
implies that a heavy ion is likely to
be able to bound multiple light atoms, especially in different vibrational
states (orbits), in which atom-atom interactions are likely to
be negligible. Such states, which can be expected to be much
more stable than the states proposed by C\^ot\'e \textit{et al.} \cite{cot02}, 
are conceptually similar to the doubly or
multiply excited states of an atom, except that the latter states have
much shorter lifetimes due to strong electron-electron interactions.
It will be interesting to find out how many atoms
can be bound in such states, and how their properties depend
on their configurations, and parameters such as the atom-ion 
mass ratio, atomic statistics, and atom-atom scattering characteristics. 

In conclusion, we have shown that in addition to the more obvious 
aspect of having a large
scattering cross section, the ``strong influence'' of 
an ion on a neighboring atom is reflected 
in the presence of a large number of bound states 
and scattering resonances around the threshold,
a situation that would normally complicates its description.
Fortunately, the very same ``strong influence'', as characterized
by a large $\beta_4$, simultaneously ensures that 
all such states follow universal properties determined solely
by the long-range potential. Further studies of few-body and many-body
systems involving ions will help to reveal their important 
roles in chemical reactions and in catalysis.

\begin{acknowledgments}
I thank Jun Ye, Alex Dalgarno, Chris Greene, Marianna Safronova,
and Hossein Sadeghpour for helpful discussions at various stages 
of this work. It was supported by NSF.
\end{acknowledgments}

\bibliography{bgao,twobody,ieatom,general}

\end{document}